\documentstyle[multicol,prb,aps,psfig]{revtex}
\draft
\tighten
\begin{document}
\title{Statistics of the Charging Spectrum of a Two-Dimensional Coulomb Glass Island.
}
\author{A.~ A.~ Koulakov, F.~ G.~ Pikus and B.~ I.~ Shklovskii}
\address{Theoretical Physics Institute, University of Minnesota,
Minneapolis, Minnesota 55455}

\date{\today}

\maketitle
\begin{abstract}

The fluctuations of capacitance of a two-dimensional island are 
studied 
in the regime of low electron concentration and strong disorder,
when electrons can be considered classical particles.
The universal capacitance
distribution is found, with the dispersion being of the order of the average.
This distribution is shown to be closely related to the shape 
of the Coulomb gap in the one-electron density of states of the
island. Behavior of the the capacitance fluctuations near
the metal - insulator transition is discussed.
\end{abstract}
\pacs{PACS numbers: 71.55.Jv}
\begin{multicols}{2}


Capacitance $C$ is conventionally understood as a well defined geometrical property
of a metallic sample.
For example, for a metallic sphere of radius $R$ capacitance $C=R$,
and when the sphere is large, in the first approximation capacitance
does not depend on the distribution of impurities
inside the sphere or on its charge. However, in very small
metallic samples fluctuations of capacitance
become observable. Recently, such fluctuations were measured
in the semiconductor quantum dots as a function of the total charge of the dot
using the Coulomb blockade phenomenon \cite{McEuen,Sivan}.
In such an experiment a small quantum dot is weakly coupled to current
leads while a gate is placed in the proximity of
the dot and is used to vary its electrostatic potential. At low
temperatures the charge of the dot is typically quantized and there 
is no significant current.
However the gate voltage can be tuned in such a way that 
the ground states with $N$ and $N+1$ electrons are degenerate.
At this gate voltage current can flow through the dot.
The resulting conductance vs. gate voltage comprises a series of sharp peaks
(charging spectrum).
The spacing between two peaks $\Delta V_{g}$ can be expressed in terms 
of the ground state energies $E_{N}$ of the dot with $N$ electrons:

\begin{equation}
e\alpha\Delta V_{g}=\Delta_{N}=E_{N+1} - 2E_{N} + E_{N -1}=e^{2}/C_{N}
\label{capacitance}
\end{equation}

Here $\alpha$ is the geometrical coefficient, $C_{N}$ is the capacitance of the dot
with $N$ electrons. This equation may be considered as a definition
of the capacitance. For a macroscopic body 
with the positive background charge $eN_0$, 
$E_{N}$ has a simple form: $E_{N}=e^{2}(N-N_{0})^{2}/2C$ 
and Eq.~(\ref{capacitance}) gives $C_{N}=C=const$. 
For the quantum dot 
the charging energy $\Delta_{N}$ was found \cite{Sivan} to have surprisingly
large relative fluctuations:
\begin{equation}
\delta \equiv \frac {(<\Delta_{N}^2> - <\Delta_{N}>^2)^{1/2}}{<\Delta_{N}>}
\sim 0.15.
\label{def_delta}
\end{equation}
Here $<...>$ denotes the averaging over $N$. Much effort has been done 
to explain such large fluctuations. First, the experimental data
were compared with the so-called constant interaction model
in which $\Delta_N = e^2/C+\eta_{N+1} - \eta_{N}$, where $\eta_N$ is the $N$-th
single-electron energy~\cite{Sivan}. The fluctuations of the spacing between 
the nearest neighbor levels are of the order of the average spacing $E_{\rm F}/N$,
where $E_{\rm F}$ is the Fermi energy. Hence,
for a dot of radius $R$
\begin{equation}
\delta \sim \frac{E_{\rm F}/N}{e^2/R} \sim \frac {r_s}R
\label{naive}
\end{equation}
where $r_s = a_B$ is the screening radius of the two-dimensional electron gas,
$a_B$ is a semiconductor Bohr radius, which is close to 10nm in GaAs.
For this value of $a_B$ and for $R \sim 200{\rm nm}$ (see Ref.~\onlinecite{Sivan}) 
Eq.~(\ref{naive}) gives fluctuations that
are substantially smaller than observed in the experiment. 
This discrepancy initiated the computer
modelling and the analytical calculations of $\delta$
(see Ref~\onlinecite{Sivan,Berkovits}).
When discussing theoretical results one should keep in mind that in all the theoretical
works $\delta$ was obtained by averaging over the different realizations of disorder,
instead of number of electrons $N$. Below we will also assume that in strongly
disordered systems there is no difference between these two definitions of fluctuations.

Analytical diagrammatic calculations
based on RPA confirm Eq.~(\ref{naive})\cite{Berkovits}. On the other hand,
the results of computer modelling \cite{Sivan,Berkovits} agree with
Eq.~(\ref{naive}) for weak interactions (large $r_s$ and $a_B$)
and lead to larger and interaction independent $\delta$  for strong interactions,
corresponding to low density electron gas. This fact was identified as a
failure of RPA in the the low density electron gas\cite{Berkovits}.
It can be interpreted easily in terms of revision of equation $r_s = a_B$
at low densities $n \ll a_B^{-2}$.
Indeed $r_s$ cannot be smaller than the average distance
between electrons $n^{-1/2}$ and at the small densities one should substitute
$ r_s =n^{-1/2}$ into Eq.~(\ref{naive}). It is not clear yet
whether such a simple modification
of  Eq.~(\ref{naive}) can quantitatively explain numerical and experimental data
\cite{Sivan}, which seem to indicate that $\delta$ is almost $R$ independent. 
Hence, it is challenging to understand what happens with $\delta (R)$
in the limit of a very low electron density.

In this paper we theoretically study the fluctuations of capacitance of 
the island in the extreme classical limit
when the quantum kinetic energy of electrons is much smaller than 
both the disorder strength and Coulomb interactions. 
We consider the case of a large
disorder when the 
ground state of the island is a Coulomb glass \cite{Efros}. Below we show that for
a piece of Coulomb glass or, in other words, a Coulomb glass island,
the fluctuations are large, $\delta$ is of the order of unity,
does not depend on $R$,
i.e. is universal for a given shape of the island. For the square 
sample we find $\delta=0.32$. (Previously a similar statement about
giant, of the order of unity, relative fluctuations of the polarizability of
the Coulomb glass island was made in Ref.\onlinecite{Baranovskii}).

We study probability density  of $\Delta_{N}$ and show that
it is a universal function of the ratio $x = \Delta_{N}/<\Delta_{N}>$.
We also discuss how the transition from the Coulomb glass to 
metal is reflected in the capacitance fluctuations.


When an electron is added to a large metallic sample its charge 
is distributed in the unique way
according to the electrostatic theory. For this reason with addition of every electron
the electric potential grows by the same amount $e^{2}/C$, or in other words $C=const$.
In the Coulomb glass the electronic states are localized, 
so that every new electron is put
into some localized site. Then electrons rearrange themselves in
the vicinity of this site. However it was shown in Ref.\onlinecite{Baranovskii} 
that such
rearrangement happens only with probability close to 1/2 on every scale. As a result
the added charge in the majority of cases
is localized in the region smaller than the size of the island. When next
electron enters the island its charge is centered near another site in the island.
The distance between this site and
the position of the previous electron fluctuates
between $0$ and $2R$. As a result the difference between energies required to bring 
two sequential electrons, $\Delta_{N}$ and capacitance  
$C_{N}$ experience roughly speaking hundred percent fluctuations.

\begin{figure}
\centerline{
\psfig{file=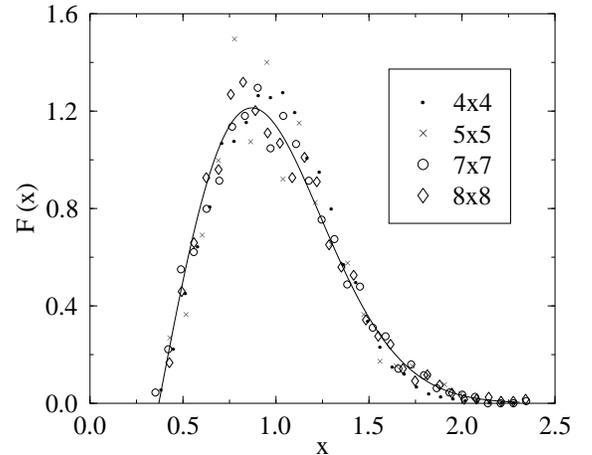,height=2.5in,bbllx=32pt,bblly=104pt,bburx=496pt,bbury=470pt}
}
\vspace{0.0in} 
\setlength{\columnwidth}{3.2in}
\centerline{\caption{
The inverse capacitance distribution is presented for 
different sample sizes.
The line is the fit by Eq. (\protect\ref{Wigner}).
\label{fig1}
}}
\vspace{-0.1in}
\end{figure}

To verify this reasoning we study the capacitance fluctuations numerically.
We use the lattice model of the Coulomb glass suggested by Efros \cite{Efros76}
and widely used to study the Coulomb gap in the density of states (DOS).
The Coulomb glass island is modelled by the square $M\times M$  lattice, 
with every site being either empty ( occupation number $n_{i}=0$ ) or occupied 
by one electron ($n_{i}=1$).
Electrons interact with each other by Coulomb interaction. The interaction
energy between the nearest lattice sites is chosen to be the unit of energy 
and the lattice constant is the unit of length.
Disorder is introduced by the random site energies $\phi_{i}$ which are 
distributed uniformly
between $-1$ and $1$. The corresponding Hamiltonian has the form
\begin{equation}
H_{\rm class} = 
\sum_{i} \left( \phi_{i} + u_i \right) 
n_{i} + \frac 12 \sum_{i,j}n_{i}n_{j}/r_{ij}
\label{Hamiltonian}
\end{equation}
Here $u_i$ is the potential due to the 
uniform background charge making the system 
electrically neutral for $N$ electrons. 
We find the ground state energies of $N-1,N,N+1$ electrons where 
$N$ is the integer part of $M^{2}/2$
and then calculate $\Delta_{N}$ using Eq.~(\ref{capacitance}).
To find the ground state we use two different methods: 
the exhaustive enumeration and the simulated annealing.
The first one is used for relatively small samples with $M \leq 5$.
We enumerate {\em all} the possible states
of $N$ electrons on $M\times M$ lattice sites
and find one with the lowest energy.
In the second method we employ the simulated annealing technique, 
running the finite-temperature classical Monte-Carlo for some time and taking the
lowest energy state. The convergence of the solution to the ground state
has been checked by doubling the time of the simulation and making sure
that the solution is not affected.
The reliable results have been obtained by this method for $M\leq 8$.
$\Delta_{N}$ has been calculated typically for 1000-2000 different 
realizations of disorder.
We have obtained $\delta = 0.32$.
Normalized distributions $F(x)$ of the ratio $x = \Delta_{N}/< \Delta_N >$
for $M=4,5,7,8$ are shown in Fig.~\ref{fig1}.
We have found $< \Delta_N > = 2.3/M$, in a good agreement with the inverse capacitance
of metallic square of the same size. 
Remarkably, $F(x)$ does not depend on $M$.
We emphasize also that 
contrary to the predictions of the constant interaction model
the inverse capacitance
of the Coulomb glass island can be both larger and {\em smaller} than that of
the metallic island of the same shape.
%
%
%
%
The fit in Fig.~\ref{fig1}
is given by the following equation:
\begin{equation}
F\left( x \right) = \left\{ 
\begin{array}{ll}
0, 				&x < x_0 \\
4(x-x_0)\exp \left( -2(x-x_0)^2 \right), 	&x \geq x_0,
\end{array}
\right.
\label{Wigner}
\end{equation}
where $x_0 = 0.37$. Two interesting features of $F\left( x \right)$
are clearly seen. Firstly, this function has a termination point
at $x = x_0$. One can easily check that $\Delta _N$ corresponding to
this point is equal to the smallest possible Coulomb interaction :
$1/r_{\rm max}$, where $r_{\rm max} = \left( M-1 \right) \sqrt{2}$
is the maximum distance between sites in the square $M \times M$. 
Secondly, $F(x+x_0)$ is identical to the Wigner surmise for the nearest 
neighbor distance distribution of the levels of a random matrix.
We shall show below that this is only an interesting coincidence.

\begin{figure}
\centerline{
\psfig{file=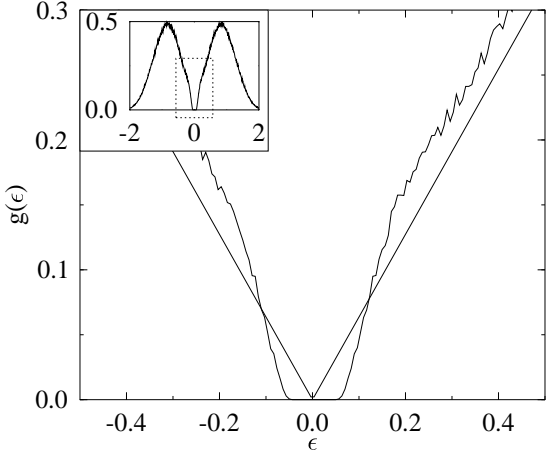,height=2.5in,bbllx=32pt,bblly=104pt,bburx=496pt,bbury=470pt}
}
\vspace{0.0in} 
\setlength{\columnwidth}{3.2in}
\centerline{\caption{DOS for the $8\protect\times 8$
sample averaged over disorder. The main picture shows the region
near the Fermi level. The Coulomb gap in the density of states
for an infinitely large sample [Eq.~({\protect \ref{Gap}})]
is presented by the straight lines.
The inset shows the general view of the DOS.
\label{fig2}
}}
\vspace{-0.1in}
\end{figure}

Now we will interpret both features establishing 
the relation between $F(x)$ and the one-electron DOS
$g(\epsilon )$
of the Coulomb glass island. The energy of the one-electron excitation
localized at $i$-th site
can be written as
\begin{equation}  
\epsilon_i = \phi_i + u_i + \sum_{j} \frac {n_j}{r_{ij}}.
\end{equation}
For an empty site, for example, it is the energy required 
to bring an electron from infinity to this site.
When averaging this DOS over different realizations
of the disorder potential or, in other words, over different samples,
we match the chemical potential in them ($\mu$ - averaging)~\cite{Efros}.
The chemical potential
of the island is situated halfway 
between the largest energy of the occupied
states and the lowest energy of the empty ones.
The corresponding DOS is shown in Figure~\ref{fig2}.
%
%
%
%
The important feature of this DOS is a linear Coulomb gap,
which at the small energies crosses over to the hard gap related to the finite
size effects ~\cite{Efros}. Linear dependence of the DOS for 
$\left| \epsilon \right| > 0.2$ 
 agrees with the analytical expression for an 
infinitely large sample
\begin{equation}
g(\epsilon) = \frac 2{\pi}\left| \epsilon \right|
\label{Gap}
\end{equation}
derived in Ref.~\onlinecite{Efros76,ES}.
The total width of the hard gap is
equal to $1/r_{\rm max}$, where $r_{\rm max}$ is the 
maximum distance between two points in the island. 
Indeed, the energy that is required to transfer an electron
from site $i$ to an empty site $j$ is equal to:
\begin{equation}
\Delta_{i\rightarrow j} = \epsilon_j - \epsilon_i - \frac 1
{\left| {\bbox r_i} - {\bbox r_j}\right|} \geq 0
\label{Delta}
\end{equation}
The minimum difference between the energies of empty and occupied
states cannot exceed the minimum interaction energy within
the island and, hence, is greater than or equal to $1/r_{\rm max}$.

Let us now explain how the one-electron DOS can be used to find $F(x)$.
Strictly speaking the
one-electron energies are not directly related to the ground state 
energies and capacitance. The excitations relevant to capacitance
are electronic polarons introduced by Efros\cite{Efros76}.
Their energies $\tilde{\epsilon_i}$ can be used to calculate 
$\Delta_N$. Indeed, $\tilde{\epsilon_i}$ is defined as the 
energy required to
bring an electron to the site $i$ and 
rearrange the other electrons in order 
to reach minimum of the total energy. 
From this definition it is obvious
that the minimum polaron energy of
the empty states is $E_{N+1}-E_N$, and the maximum polaron energy of 
the occupied states is $E_N-E_{N-1}$. It is clear that $\Delta_N$ is
just the energy gap between these two polaron states. 
It is known, however, that in two dimensions the polaron energies 
are very close to the one-electron ones~\cite{Efros}. This means
that $\Delta_N$ can be well approximated by the 
energy difference between the lowest empty and the highest occupied 
{\em one - electron} states. This immediately
explains the existence of the termination point of $F(x)$:
$\Delta_N$ cannot be smaller than the smallest interaction between 
two electrons within the island. Moreover the function $F(x)$
can be related to the one-electron DOS
$g(\epsilon )$ in the following way.

As the inverse capacitance of an island is equal to the difference
between the lowest empty and the highest occupied states' energies,
our problem is to find the distribution function of this
difference. The probability 
$P\left( \Delta \geq \epsilon \right)$
to have it bigger or equal 
than certain value $\epsilon$ is equal to the
probability not to find energy levels
in the region $-\frac {\epsilon}2 < \epsilon ' < \frac {\epsilon}2$.
Assuming the Poissonian statistics of the level distribution
we arrive at:
\begin{equation}
P\left( \Delta \geq \epsilon \right) =
\exp \left( - M^2 \int_{-\frac{\epsilon}2}^{\frac{\epsilon}2}g(\epsilon ') 
d\epsilon '
\right).
\label{Prob}
\end{equation}
Here $g(\epsilon)$ is assumed to be normalized to unity: 
$\int_{-\infty}^{\infty}g(\epsilon ') d\epsilon ' = 1$.
The expression in the exponential is the average number of electrons in 
the band of energies $\left( -\frac {\epsilon}2, \frac {\epsilon}2\right)$.
The probability density of $\Delta$ is, hence, equal to
\begin{equation}
\begin{array}{ll}
F(\epsilon) &{\displaystyle  
=  - \frac {dP(\Delta \geq \epsilon)}{d\epsilon} }\\
&{\displaystyle  = 2N g\left( \frac {\epsilon}2 \right)
\exp \left( - 2 M^2 \int_0^{\frac{\epsilon}2}g(\epsilon ') 
d\epsilon ' \right)} 
\label{Dens}
\end{array}
\end{equation}
This expression establishes the general relationship
between the one-electron DOS and the 
inverse capacitance distribution function for the Coulomb
glass island. Having been applied to the numerically
obtained DOS it gives a very good agreement with the
actual $F(\epsilon)$. Substituting Eq.~(\ref{Gap}) into Eq.~
(\ref{Dens}) one arrives at the Gaussian asymptotic behavior of
$F(x)$: $\ln F(x) \propto  - x^2$, $1 \ll x \ll M$.
This asymptotic, however, does not persist in three dimensions, where 
the DOS has a different form.

\begin{figure}
\centerline{
\psfig{file=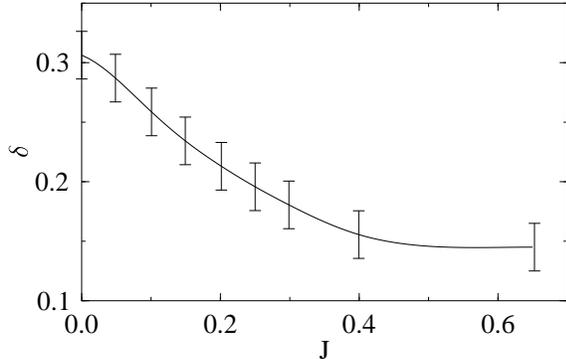,height=1.9in,bbllx=72pt,bblly=101pt,bburx=490pt,bbury=368pt}
}
\vspace{0.0in} 
\setlength{\columnwidth}{3.2in}
\centerline{\caption{
The relative inverse capacitance fluctuations of $4 \times 4$
island as a function of the hopping matrix element $J$. 
\label{fig3}
}}
\vspace{-0.1in}
\end{figure}

\begin{figure}
\centerline{
\psfig{file=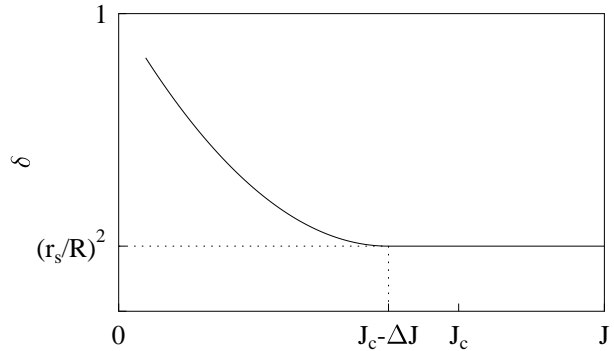,height=1.9in,bbllx=72pt,bblly=101pt,bburx=490pt,bbury=368pt}
}
\vspace{0.0in} 
\setlength{\columnwidth}{3.2in}
\centerline{\caption{The relative inverse capacitance fluctuations 
of the three-dimensional island as given by Eq.~({\protect \ref{naive1}})
for ${\protect J < J_c - \Delta J}$ and by Eq~({\protect \ref{naive2}}) 
for ${\protect J > J_c - \Delta J}$.
\label{fig4}
}}
\vspace{-0.1in}
\end{figure}

Let us now examine what happens with $\delta$ when a hopping term 
is added to the classical Hamiltonian, 
given by Eq.~(\ref{Hamiltonian}):
\begin{equation}
H = H_{\rm class}  - J\sum_{\left< i, j \right>} a^{\dagger}_i a_j.
\label{QH}
\end{equation}
Here $J$ is the hopping matrix element and $a^{\dagger}_i$
is the creation operator of the electron at site $i$.
The summation is carried over the neighboring sites.
The ground states of this Hamiltonian were found numerically using
the Lanczos algorithm. The system considered was of size $4 \times 4$
lattice sites, with up to $7$, $8$ and $9$ electrons. 
The results are depicted in Fig.~\ref{fig3}.
%
%
%
%
As it is seen that $\delta$ decreases by a factor of $2$ from $J=0$
to $J=0.4$. 
Such a decay is consistent
with the tendency of metallic samples to have smaller
capacitance fluctuations.
At the same time the shape of $F(x)$ becomes more symmetric in 
agreement with Ref.~\onlinecite{Sivan,Berkovits}. 
Unfortunately we were not able to do such calculations for $M > 4$.
Hence, the existing numerical data leave open the challenging question of
how the crossover happens between $\delta\sim 1$ in the classical case and
Eq.~(\ref{naive}) in the quantum one for large enough samples.
Below we try to answer this question
concentrating on the three-dimensional case where the  insulator-metal  transition
happens at some critical value $J=J_c$.
We assume that this transition is accompanied by the divergency of
the wavefunction correlation length $\xi$ and
the dielectric constant $\kappa$: $\xi=|J_c-J|^{-\nu}$ and $\kappa=|J_c-J|^{-\zeta}$.
It was argued that $\kappa \sim (\xi/r_s)^2$, where $r_s$ is
the screening radius of the three-dimensional degenerate Fermi-gas, 
or  $\zeta=2\nu$ ( see Ref.~\onlinecite{Gefen} ).
When $J$ approaches $J_c$ from the insulator side
the growth of $\kappa$ plays a very important role in the distribution of charge of
the added electron,
even when $\xi$ is still much smaller than a sample size $R$.
It is well known that if a localized charge is put inside a dielectric sample with
$\kappa\gg 1$ the sample becomes polarized in such a way that almost
all of the added charge appears on its surface in the form of induced charge.
Only a small fraction of charge $e/\kappa$ remains localized inside.
If the second electron is added
to the sample at the distance $r$ from the first 
the interaction energy between these two consecutively
added electrons fluctuates by
$e^2/\kappa R \ll e^2/ R$, provided that $r$ fluctuates in the
range $0<r<2R$ for a sphere of radius $R$. 
Therefore
\begin{equation}
\delta \sim \ 1/\kappa \sim (r_s /\xi)^2 \sim (J_c-J)^{2\nu}
\label{naive1}
\end{equation}
Eq.~(\ref{naive1}) is valid only if $\xi\ll R$, or $J < J_c - \Delta J$,
where $\Delta J = \left( r_s / R \right)^{1/\nu}$
. At  $\xi= R$
the relative capacitance fluctuations saturate at 
\begin{equation}
\delta  \sim (r_s /R)^2
\label{naive2}
\end{equation}
Eq.~(\ref{naive2}) is the three-dimensional analog of  Eq.~(\ref{naive}). 
It can also be
obtained from the assumption that fluctuations of $\Delta_N$ are equal to the
fluctuations of the spacing between the one-electron quantum levels.
Predicted behavior of function $\delta(J)$ is schematically depicted in
Fig.~\ref{fig4}.

%
%
%
%

We conjecture that in the two-dimensional case, where as it is commonly believed $\xi$
grows monotonically with $J$,
the crossover from $\delta \sim 1$ to Eq.~(\ref{naive}) happens in the similar way:
$\delta \sim \ r_s /\xi$ at $ r_s \ll \xi\ll R$
and $\delta  \sim r_s /R$ at $ \xi\gg R$. Schematically this behavior
is similar to the one shown in Fig.~\ref{fig4}.
Our numerical results do not contradict to this prediction.

Up to now we have dealt with the unscreened Coulomb interaction between electrons.
In case if a metallic gate is situated at a small distance $d$ from the plane
of the island the long-range part of the Coulomb interaction is screened.
Let us now discuss how this screening affects our results. 
Consider first the case of an extremely close gate when one can 
completely ignore the interactions between electrons. In this case the 
constant (zero) interaction model is valid, $N$ electrons occupy $N$ lowest
one-electron levels $\eta_k$, and $\Delta_N = \eta_{N+1}-\eta_{N}$ is just
a nearest neighbor level spacing.
At $J=0$ it has the Poisson distribution, therefore $\delta=1$.
In the metallic phase (large $J$) the
random matrix theory is applicable and $ \delta=0.52 $ (see Ref.~\onlinecite{Sivan}).
We studied this crossover for up to $10^4$ realizations of disorder 
in the square $4\times 4$,
with the Coulomb interaction replaced by $1/r - 1/(r^2 + (2d)^2)^{1/2}$, $d=1/2$. 
We obtained $\delta = 0.67$ at $J=0$ and $\delta = 0.39$  at $J=0.4$ in
the qualitative agreement with our expectations.
At $J=0$ we also observed a drastic change in the form of $F(x)$: 
the gap at small positive $x$ disappears and $F(x)$
approaches the Poisson distribution.
The most remarkable feature of the observed distribution is that 
there exists a very small tail of $F(x)$ at $x<0$,
so that the differential capacitance of a strongly screened disordered island 
can be {\em negative}.
The possibility of the negative capacitance in the presence of a gate
was recently demonstrated in Ref.\onlinecite{Raikh}. The observed probability 
to have negative capacitance is consistent with the calculations of these authors.
For our set of parameters it is equal to $3\cdot10^{-4}$. 
 
In conclusion, we have found the large universal relative fluctuations of 
capacitance of the Coulomb glass island and described 
a scenario of
their decay when the system undergoes the insulator-metal transition.
We are grateful to  M.\ M.\ Fogler for valuable discussions.
This work was supported by the NSF Grant No.\ DMR-9321417
and by University of Minnesota Supercomputer Institute.
\vspace{-0.2in}

\end{multicols}
\end{document}